# Excitation laser energy dependence of the gap-mode TERS spectra of $WS_2$ and $MoS_2$ on silver.

Andrey Krayev[1*#], Eleonora Isotta[2*#], Lauren Hoang[3], Jerry A. Yang[3], Kathryn Neilson[3], Minyuan Wang[4], Noah Haughn[4], Eric Pop[3,5], Andrew Mannix[5], Oluwaseyi Balogun[6,7] and Chih-Feng Wang[8].

1. HORIBA Scientific
2. Department of Materials Science and Engineering, Northwestern University
3. Department of Electrical Engineering, Stanford University
4. UC Davis, Department of Chemistry
5. Department of Materials Science and Engineering, Stanford University
6. Department of Mechanical Engineering, Northwestern University
7. Department of Civil and Environmental Engineering, Northwestern University
8. Pacific Northwest National Laboratory

*-Authors contributed equally
#-Corresponding authors

## ABSTRACT

In this work we present a systematic study of the dependence of the gap mode tip-enhanced Raman scattering (TERS) response of the mono- and bi-layer $WS_2$ and $MoS_2$ on silver as a function of the excitation laser energy in a broad spectral range from 473nm to 830 nm. For this purpose, we collected consecutive TERS maps of the same area in the sample containing mono-and bi-layer regions with the same TERS probe with 6 different excitation lasers. To decrease the number of collected TERS maps, we used for the first time to the best of our knowledge, concurrent excitation and collection with two lasers simultaneously. We found that the $E_{2g}/A_{1g}$ peak intensity ratio for the bilayer $WS_2$@Ag and the ratio of the A'/$A_{1g}$ peak intensity of the out-of-plane mode for the mono- and the bilayer, change in a significantly non-monotonous way as the excitation laser energy is swept from 1.58 eV to 2.62 eV. The former ratio increases at energies corresponding to A and B excitons (~2.0 eV and 2.4 eV correspondingly) in bilayer $WS_2$. The absolute intensity of the A' peak in the monolayer, and correspondingly the A'/$A_{1g}$ ratio, is surprisingly high at lower excitation energies, but dips dramatically at the energy corresponding to the A exciton, being restored partially in between A and B excitons, but still showing the descending trend as the excitation laser energy increases.

A somewhat similar picture was observed in mono- and bi-layers of $MoS_2$@Ag, though the existing set of excitation lasers did not match the excitonic profile of this material as nicely as for the case of $WS_2$.

We attribute the observed behavior to the presence of intermediate (Fano resonance) or strong (Rabi splitting) coupling between the excitons in transition metal dichalcogenides (TMDs) and the plasmons in the tip-substrate nanocavity. This is akin to the so-called Fano (Rabi) transparency experimentally observed in far field scattering from transition metal dichalcogenides between two plasmonic metals.

The possibility of the formation of intermediate/strong coupling between the excitonic resonances in TMDs and the nanocavity re-evaluates the role of various resonances in the gap-mode TERS and should become an important factor to be considered by TERS practitioners during planning the experiments.

Finally, based on observed phenomena and its explanation, we propose the "ideal" substrate for efficient TERS and tip enhanced photoluminescence (TEPL) measurements.

## INTRODUCTION.

Raman spectroscopy proved to be an extremely useful technique for the characterization of 2D materials like graphene or transition metal dichalcogenides (TMDs). In TMDs many optoelectronic properties, including resonant Raman response, are mediated by excitons which happily survive at room temperature. As a result, Raman spectra of TMDs may change quite dramatically as a function of the energy of the excitation laser when a certain excitonic resonance is excited. Moreover, different Raman modes may be mediated by different excitons and therefore, exhibit different Raman excitation profiles[1,2].

The importance of understanding the resonant Raman response in TMDs and their Raman excitation profiles in the NIR-to-UV spectral range was recognized quite early, and a number of papers have been published on this subject[1–6].

Tip enhanced Raman scattering (TERS) imaging, in particular in its gap mode implementation when a thin sample is sandwiched between the TERS probe and a plasmonic or just well reflecting metallic substrate [7], is getting increasingly popular for the characterization of 2D materials in general and TMDs in particular. The strong enhancement of the Raman response in TERS combined with a significantly improved spatial resolution in the order of 10-20nm, which can be routinely achieved in TERS maps in ambient conditions, make TERS imaging a technique of choice for the spectroscopic nanometer-scale characterization of lateral [8–10] and vertical [11–13] TMD heterostructures, spatially inhomogeneous doping in exfoliated layers[14,15], strain related phenomena in nanobubbles of TMDs [16,17], growth and aging-related defects [18], Moire patterns in twisted bilayers [19]. An additional advantage of TERS imaging is that the TERS maps can be cross-correlated with other physical characteristics provided by scanning probe microscopy like topography, surface potential, photocurrent etc. which may be very beneficial for the interpretation of the data collected.

Up to date, the majority of TERS experiments are carried out with red lasers, 632.8 nm or 638 nm due to the preferential use of gold coated or etched gold TERS probes. At the same time, the importance of varied excitation in TERS measurements was recognized [13], when the use of just one excitation wavelength could lead to wrong conclusions regarding the composition of $WS_2$-$WSe_2$ vertical heterostructures.

In this work we conducted for the first time a systematic study of the gap-mode TERS response of $WS_2$ and $MoS_2$ mono- and bi-layers in intimate contact with silver as a function of the excitation laser wavelength in 473-830nm spectral range. Protected silver probes and silver substrate were chosen in order to extend the TERS study to shorter wavelengths ( λ<580nm ) which is not achievable with gold due to the strong interband transition absorption below about 580 nm.

## METHODS.

Monolayer and bilayer $WS_2$ were directly grown on 100 nm $SiO_2$/Si substrates by hybrid metal organic chemical vapor deposition (CVD) using diethyl sulfide (DES) and ammonium metatungstate (AMT) precursors, as detailed in a previous work.[20] Briefly, 0.6 g AMT and 0.1 g potassium hydroxide are dissolved in 30 ml deionized water and dip-coated on the edges of the substrate. The substrate is then annealed in the furnace at 775°C for 9 h with a DES flow rate of 0.1 sccm. $N_2$ and $H_2$ are flowed as carrier gases during the growth.

$MoS_2$ was grown on 1.5x1.5 cm chips of 90 nm thermal SiO2/Si wafers by solid-source CVD as described in Smithe et al.[21] The $SiO_2$/Si wafers were treated with a hexamethyldisilazane (HMDS) vapor singe and prime. Immediately prior to growth, the chips were cleaved from the full wafer, and several 25 uL drops of a perylene-3,4,9,10 tetracarboxylic acid tetrapotassium salt (PTAS) seed promoter were placed around the chip perimeter. The chips were then placed into a 50 mm tube furnace with the solid-source S and $MoO_3$ precursors. The growth conditions were 750°C, 800 Torr, 5 min in 30 sccm Ar ambient. For both $WS_2$ and $MoS_2$ samples, blanket Ag/Au (70/50 nm) was electron-beam deposited in a Kurt J. Lesker high vacuum evaporator (LAB 18 system). The evaporation was done at a base pressure of $\sim 1\times 10^{-7}$ Torr at an evaporation rate of 1 Å/s.

TMD crystals were stripped from the growth substrate following the procedure proposed earlier[22]. As a result we had corresponding $WS_2/MoS_2$ crystals in intimate contact with silver, but flipped "belly up".

AFM and TERS measurements were conducted on a LabRam-Nano AFM-Raman system (HORIBA Scientific) that was modified to inject additional lasers in between the Raman and the AFM parts to allow concurrent excitation/collection with 2 lasers simultaneously. A detailed sketch of the setup is provided in Figure S1in the supporting information section. Excitation and collection of TERS signal was done using the side 100X, 0.7 NA objective (Mitutoyo) inclined at 25º to the sample plane. Type II protected silver probes (HORIBA Scientific) based on Access-SNC AFM cantilevers (APPNano) were used for both the SPM and TERS characterization. Laser power past the objective for individual wavelengths was kept at the level of 100-320 µW.

# RESULTS and DISCUSSION.

**Excitation wavelength dependence of TERS response of WS$_2$@Ag.**

To access the difference in TERS response in mono- and bilayer WS$_2$@Ag, TERS maps of the same area were collected using the same TERS-active probe with 473nm (2.62 eV), 532nm (2.33 eV), 594nm (2.09 eV), 632.8 nm (1.96 eV), 671 nm (1.85 eV) and 785 nm (1.58 eV) lasers. In order to reduce the number of TERS maps collected, 632.8nm -785nm and 532nm-594nm lasers were used concurrently, which provided a unique possibility to collect on- and off-resonant TERS spectra in a single map. The use of low density grating (150 groves/mm) was mandatory to provide sufficient spectral range that would include the Raman bands at both excitation wavelengths for corresponding laser pairs. We investigated three different monolayer-bilayer crystals in two different samples to ensure that the results are sufficiently reproducible.

Despite the topography of the samples being relatively flat, as should be expected from the sample preparation procedure, the presence of thicker layer islands in the crystals was clearly visible in the contact potential difference (CPD) images, as well as in the topography itself, mostly due to the clearly distinguishable borderline between the mono- and bi- (or higher) layer areas (Figure 1).

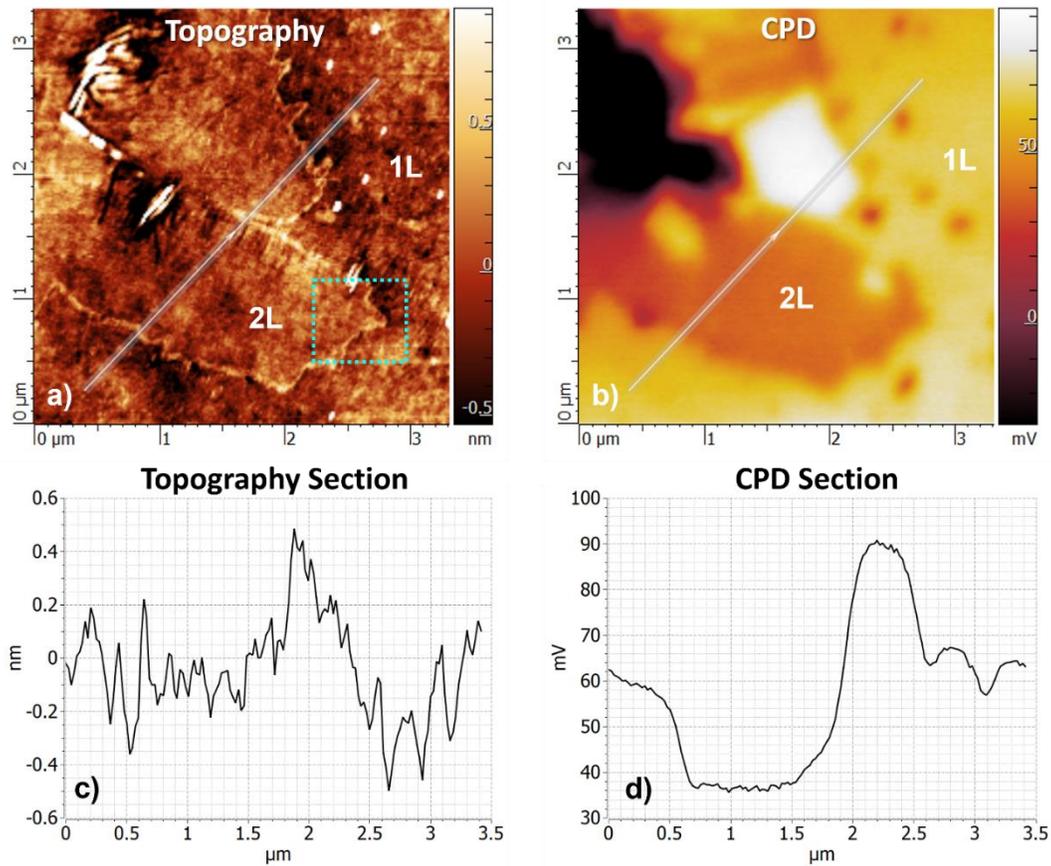

**Figure 1.** a) AFM topography and b)-corresponding contact potential difference (CPD) images of the bilayer/monolayer WS$_2$@Ag, crystal #3. c), d)- corresponding topography and CPD section graphs. Bilayer-monolayer junction line is clearly distinguishable in the topography image, while the CPD image clearly shows the bilayer region. TERS maps with varied excitation were collected over the area within the dotted cyan rectangle in panel a).

TERS spectra of the mono- and the bilayers of WS$_2$ were dominated by E'(1L)/E$_{2g}$(2L) in-plane and A'(1L)/A$_{1g}$(2L) out-of-plane peaks. In the monolayer strongly interacting with the metallic substrate, A' becomes A$_1$ due to the change of the symmetry group [23,24] which splits in two peaks. For clarity we'll call the lower energy one at ~410 cm$^{-1}$ -A', and the higher energy one at ~428 cm$^{-1}$ - A$_1$ (Figure 2 e,g).

TERS maps and corresponding TERS spectra averaged over the bilayer (blue spectra) and the monolayer (red spectra) for crystal #3, the excitation profile of which was closest to the average line, are shown in Figure 2. TERS maps and corresponding spectra for crystal #1 and #2 are shown in Figure S2 and Figure S3 in the supporting information section.

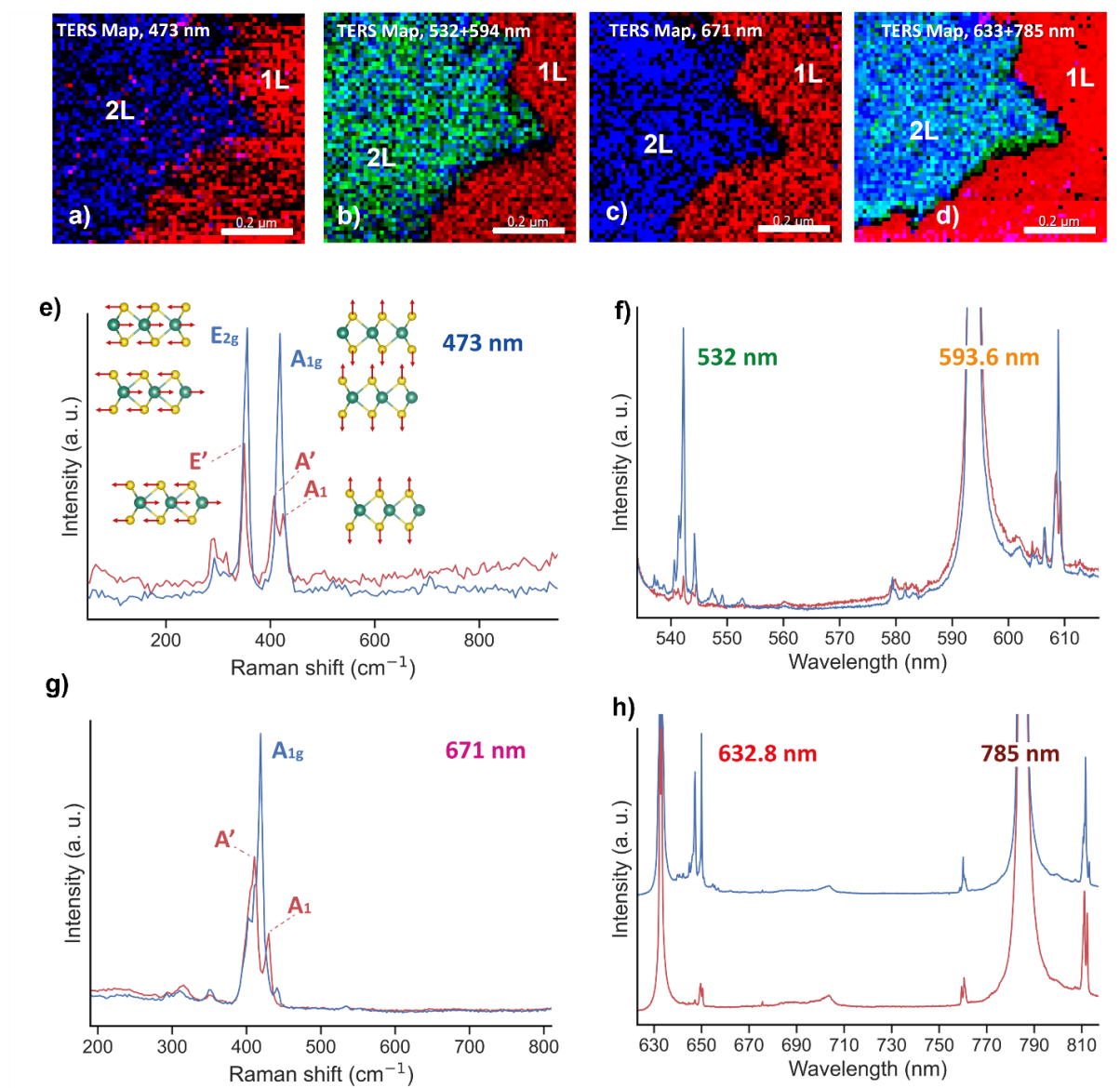

**Figure 2. a)-d)** Combined TERS maps collected using the same TERS probe over the same area of WS$_2$@Ag crystal containing the bi- and the monolayer regions with a) 473nm, b) concurrent 532 nm and 594 nm, c) 671 nm, and d) concurrent 632.8 and 785 nm excitation. Blue and red colors in panels (a-c) represent the intensity of A$_{1g}$ and A$_1$ peaks, respectively. In panels (b,d) blue and red colors represent the intensity of A$_{1g}$ and A$_1$ peaks

at 594 nm and 785 nm excitations, respectively, while green color in the same panels represents the intensity of the $A_{1g}$ mode at 532 nm and 633 nm, respectively. Scale bar in (a-d) is 200 nm. e) through h) TERS spectra averaged over the bilayer (blue spectrum) and the monolayer (red spectrum) at corresponding excitation wavelengths. Please note that for the concurrent excitation with two lasers, the x axis is in nm as we can not normalize the Raman shift for both lasers simultaneously.

Thanks to the use of 150 grooves/mm grating, we had enough spectral range to record the Raman peaks for both 785 nm/632.8 nm and 594 nm/532 nm laser pairs as clearly visible in Figure 2 f) and h). TERS spectra averaged over the bi- and monolayers and intensity-normalized to the height of the $A_{1g}$ peak in the bilayer are presented in Figure 3. This time, for every excitation wavelength, the TERS spectra were properly calibrated in wavenumbers for proper comparison.

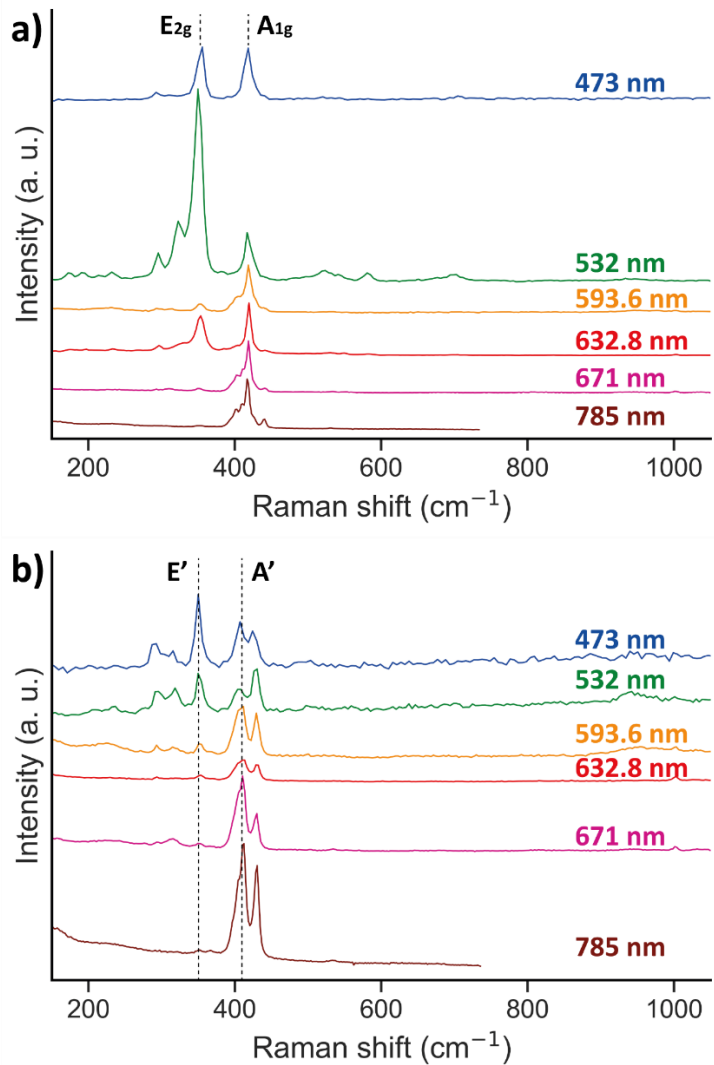

Figure 3. a) TERS spectra averaged over the bilayer $WS_2$@Ag with different excitation lasers and normalized by the height of $A_{1g}$ peak. Spectra were vertically offset for clarity; b) TERS spectra averaged over the monolayer $WS_2$@Ag normalized with the same coefficients as the bilayer spectra in panel a). Please note a monolayer-like shoulder to the left of $A_{1g}$ peak in TERS spectra of the bilayer at 785 nm, 671 nm and 594 nm excitation.

As we can see from Figure 3-a), in the bilayer region the relative height of $E_{2g}$ peak as compared to the height of $A_{1g}$ peak was changing non-monotonously. $E_{2g}$ peak was practically invisible with 785 nm and 671 nm excitation ( below the energy of A exciton in the bilayer $WS_2$), it became very much comparable with $A_{1g}$ peak at 632.8 nm excitation (very close to A exciton), went down at 594 nm excitation (in between A and B excitons), shot up dramatically at 532 nm (at B exciton) and then became more-or less equal to $A_{1g}$ peak again at 473 nm.

The excitation dependence of TERS spectra of the monolayer $WS_2$ was quite significantly different. The first striking difference was the red shift of A' peak as compared to $A_{1g}$, something we should expect taking into account the intimate contact of the monolayer with metallic silver substrate and, correspondingly, expected doping of $WS_2$ monolayer by the silver substrate [15]. Second, the A' peak appeared as a doublet, the higher-energy component of which could be assigned to the $A_1$ mode of a monolayer strongly interacting with the metallic substrate [23]. It's important to note that the absolute intensity of the A' peak (the monolayer spectra were intensity-normalized with the same coefficients as the bilayer spectra) was highest at 785 nm excitation, it dropped down dramatically at 632.8 nm, somewhat restored at 594 nm, went down at 532 nm and up again at 473 nm. Rather surprisingly, E' peak height was rising practically monotonously with the excitation laser energy, very much unlike its $E_{2g}$ analog in the bilayer.

Coming back to the TERS spectra of the bilayer, we should note that at the left-hand side of the $A_{1g}$ peak at 785 nm, 671 nm and 594 nm excitation wavelength there was a significant shoulder that looked very much like the monolayer response. We can thus state that the TERS response of the bilayer was comprised of the TERS response of the bottom layer that was strongly interacting with silver and showed a monolayer-like behavior and the top layer decoupled from the metallic substrate that dominated the spectra intensity-wise. Further down the manuscript we'll show additional evidence for such an interpretation.

In order to make the above description of the evolution of the TERS spectra more intuitive and easier to understand, we plot the graphs showing the dependence of the $E_{2g}/A_{1g}$ peak height ratio for the bilayer, E'/A' peak ratio for the monolayer and the A'/$A_{1g}$ peak ratio as the function of the excitation energy (Figure 4).

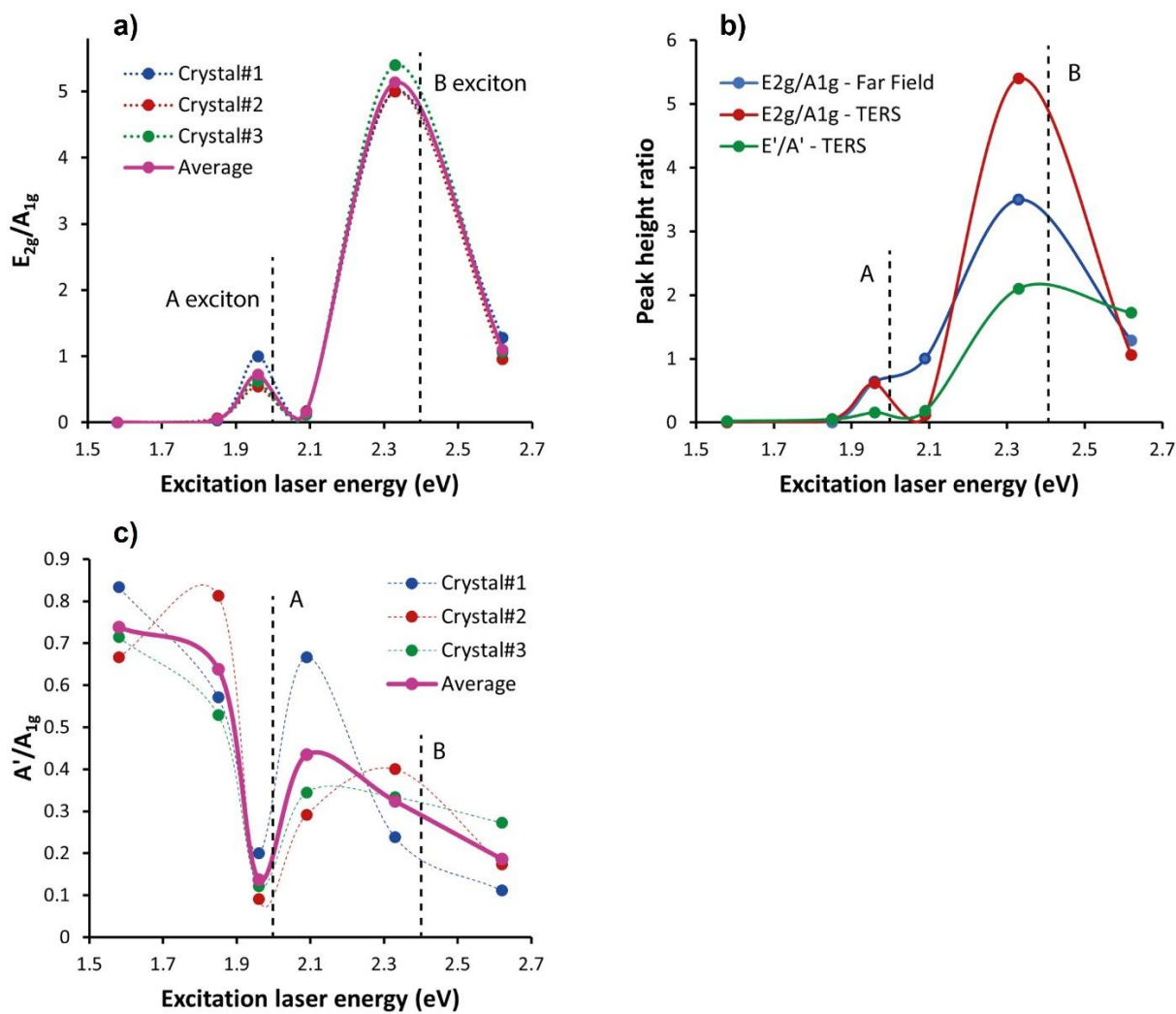

**Figure 4.** a) $E_{2g}/A_{1g}$ peak height ratio as a function of the excitation laser energy for the bilayer $WS_2$@Ag. b) TERS $E_{2g}/A_{1g}$ ratio as a function of the excitation laser energy for crystal #3 (red line) same ratio for the far-filed spectra (blue line) and TERS E'/A' peak height ratio as a function of the excitation laser energy for the monolayer $WS_2$@Ag (green line). For the far-field curve $E_{2g}/A_{1g}$ ratio, the 785 nm (1.58 eV) was assigned to zero as the measurable far field signal was absent at this wavelength. c) TERS $A'/A_{1g}$ peak height ratio as a function of the excitation laser energy. Please note that the deep dive at around 2 eV corresponds to the A exciton position in $WS_2$. Connecting lines are for eye guidance only.

The use of peak height ratios for corresponding excitation laser wavelengths/energies instead of the absolute peak heights *de facto* cancels out the influence of the instrumentation throughput function which is wavelength-dependent (this includes varying sensitivity of the EM-CCD, parasitic wavelength-dependent light absorption in the objective lens, variable reflection efficiency of the gratings etc).

Looking at the graph in Figure 4-a), we see that the $E_{2g}/A_{1g}$ ratio in TERS spectra of the bilayer goes up at energies corresponding to A and B excitons in bilayer $WS_2$ [25]. Comparing the TERS and the far-field (collected when the tip was engaged with the sample in regular semi-contact mode with average tip-sample distance of about 20nm) operation dependence of the $E_{2g}/A_{1g}$ ratio (Fig.4-b) we see that to a great extent the two curves repeat each other, which indicates that the peak ratio variation is not related to peculiarities of the gap mode TERS response. This should be expected as the spectrum of the bilayer is dominated by the contribution of the second layer which is screened from the silver substrate by the first layer. The E'/A' ratio in the TERS response of the monolayer $WS_2$ did not show any significant increase at A exciton. As was described earlier, the rise of the intensity of E' peak in TERS spectra of the monolayer was rather monotonous.

Probably the most interesting and consequential for the TERS practitioners is the $A'/A_{1g}$ graph in Figure 4-c). A deep dive of the relative intensity of the A' TERS peak at A exciton energy was very consistent, across all the crystals probed, as well as the descending trend for this ratio as the excitation laser energy was increasing. This brings us to a rather counterintuitive conclusion that for successful TERS imaging of the monolayer TMDs in intimate contact with a plasmonic metallic substrate one should keep the excitation laser energy below the A exciton, which again- rather counterintuitively- makes broadly available 785nm laser the best choice for such measurements. We should also note that silver still exhibits strong plasmonic properties at around 800nm due to the sharp rise of the modulus of negative real dielectric constant and relatively slowly changing imaginary part [26].

Further, based on the consistently strong TERS response of the bilayer *de facto* decoupled from underlying silver, we propose single layer h-BN-capped gold or silver films as the "ideal" substrate for the gap-mode TERS and tip-enhanced photoluminescence (TEPL) measurements. This substrate would simultaneously preserve the enhancement in the tip-substrate nanocavity and decouple the sample from the metallic substrate. This substrate is very much in line with the silica- or alumina-capped gold substrates proposed and tested previously[27–29]. Single layer (1L) h-BN capping provides several important advantages over the atomic layer deposition-grown silica films: first, it provides consistently thin uniform 3-4 angstroms dielectric layer, which should give much stronger enhancement compared to 2-3 nm thick silica layer. In addition, 1L h-BN enables an atomically smooth surface that benefits consequent exfoliation of 2D materials or the deposition of other research samples. Recent advances with the centimeter-scale synthesis of h-BN monolayer on Ge monocrystalline wafers [30] provides a straightforward route for fabrication of these prospective substrates when a freshly grown monolayer h-BN is capped with gold or silver and can be stripped on demand following the procedure used in this work [22].

**Excitation wavelength dependence of TERS response of $MoS_2$@Ag.**

The evolution of the TERS response of $MoS_2$@Ag samples produced with an identical procedure to their $WS_2$ counterparts was investigated in 830nm-532nm spectral range. Rather unexpectedly, we did not observe any measurable TERS enhancement in the mono- or bi-layer $MoS_2$@Ag with 473nm excitation, even though the same TERS probe produced reasonable TERS signal for

WS$_2$@Ag samples. Before we describe the results obtained, a few important notes should be made. First, A and B excitons in MoS$_2$ are closer to each other as compared to the case of WS$_2$, at ca 1.88 eV and 2 eV, respectively [31]. The existing set of CW lasers in our setup did not allow to properly track the range next to- and in-between A and B excitons. In addition, E$_{2g}$/E' and A$_{1g}$/A' are closer to each other in MoS$_2$ compared to WS$_2$, so we had to use higher density gratings to properly resolve the bands. But even with these limitations, we did not have to run all six maps with individual single-wavelength excitations; 632.8nm and 594nm excitation lasers were used concurrently.

Looking at the normalized TERS spectra of the bi- and monolayer MoS$_2$ (Figure 5), we see several interesting spectral peculiarities that were absent in WS$_2$.

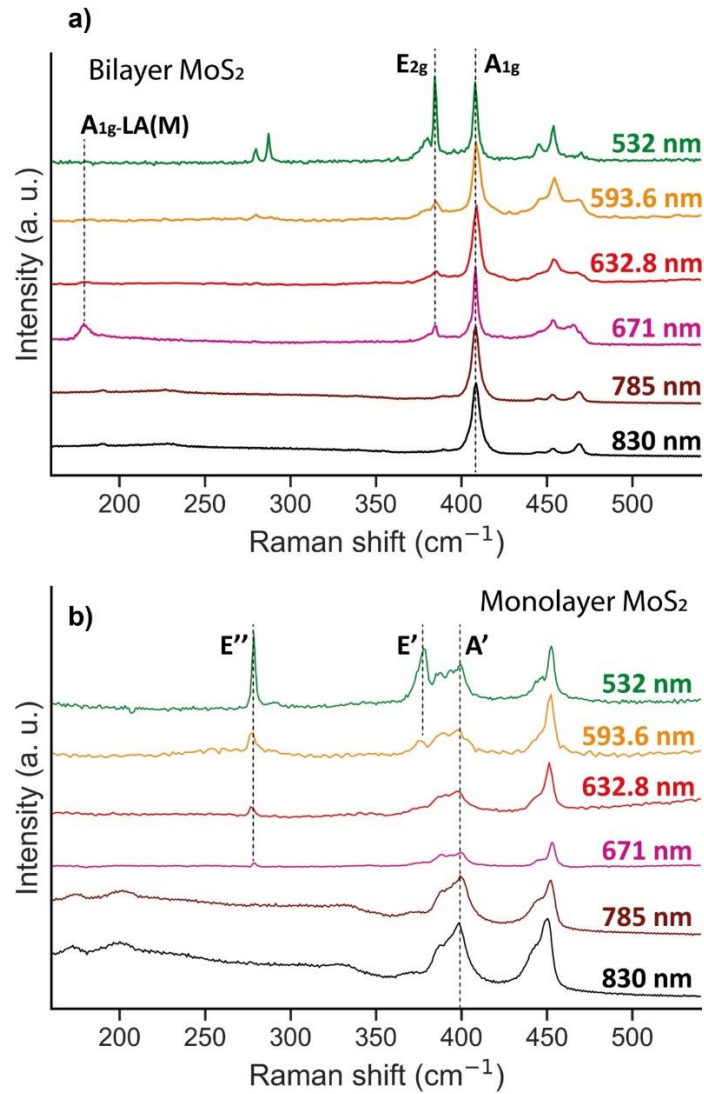

**Figure 5. a) Averaged TERS spectra of the bilayer MoS$_2$ on silver normalized by the height of A$_{1g}$ peak collected with 6 excitation laser wavelengths in 532nm-830nm spectral range. Please note the sharp rise of A$_{1g}$-LA(M) peak at 671nm excitation. b) Averaged TERS spectra of the monolayer MoS$_2$ on silver normalized with the same coefficients as the corresponding TERS spectra of the bilayer.**

First, in the monolayer spectra starting from 671nm excitation we observed a peak at ca. 277cm$^{-1}$ which we can tentatively assign to a rarely observed E" mode which is Raman active, but silent in back-scattering geometry. Quite unexpectedly, this band became the most intense band in the TERS spectra at 532nm excitation. In the bilayer region with the same 532nm excitation, next to the 277cm$^{-1}$ band we observed another band at ca 286cm$^{-1}$, which can be assigned to $E_{1g}$, the two-layer/bulk counterpart of E"[32]. E"/$E_{1g}$ is a doubly degenerate mode [32], and this may explain the splitting in TERS spectra of the bilayer, where the first layer directly contacting silver and the second one screened from the metallic substrate are not equivalent.

Despite the complex A' peak -with at least two additional peaks forming a shoulder on the lower energy side- was somewhat similar to the A' peak in $WS_2$, we did not observe the 1-L like shoulder next to the $A_{1g}$ peak in the bilayer TERS spectra, like we did in the case of $WS_2$. Another interesting feature observed in the TERS spectra of bilayer $MoS_2$ was a sharp rise at 671 nm excitation of another rarely observed peak at ca. 176cm$^{-1}$ which was assigned to $A_{1g}$-LA(M)[33].

Looking at the $E_{2g}/A_{1g}$ and A'/$A_{1g}$ peak ratio (Figure 6), we can see a general similarity with the results obtained on $WS_2$.

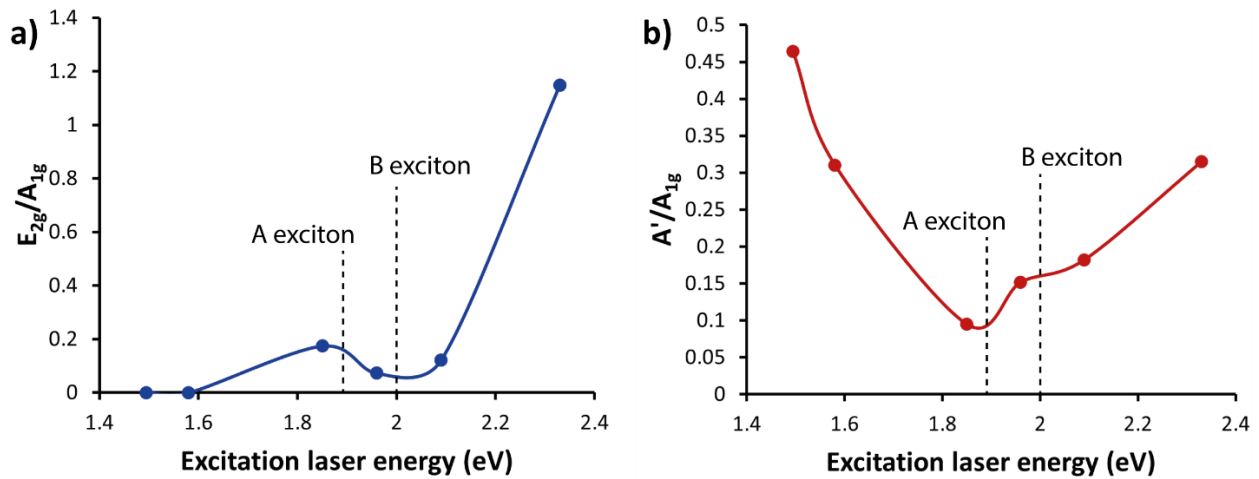

**Figure 6. a) $E_{2g}/A_{1g}$ TERS peaks ratio for the bilayer $MoS_2$ on silver as a function of the excitation laser energy. b) A'/$A_{1g}$ TERS peaks ratio as a function of the excitation laser energy. Similar to the case of $WS_2$, this ratio significantly decreases at the energy corresponding to A exciton, while reaching the maximum value at the lowest excitation laser energy. Connecting lines are for eye guidance only.**

The $E_{2g}$ mode became noticeable first time only in the vicinity of the A exciton energy, whereas it was unobservable at 1.49 eV (830nm) and 1.58 eV (785nm). Similarly, the A'/$A_{1g}$ peak height ratio dived down at the same energy close to the A exciton, though unlike in the case of $WS_2$, the A'/$A_{1g}$ ratio kept rising as we moved from 1.85 eV to 1.96, 2.09 and 2.33 eV.

Summarizing the observations made in TERS spectra of the mono- and bi-layer $MoS_2$@Ag, we can say that similarly to the case of $WS_2$, TERS spectra of the monolayer $MoS_2$ showed stronger response at the excitation laser energy below A exciton. In addition, several resonant peaks appeared in the spectra of the mono- and bilayers at different excitation laser energies, which highlights the fascinatingly complex nature of the electron-phonon/exciton-phonon coupling in TMDs and reminds us that the definition of resonant Raman conditions in these 2D semiconductors should not be given without specifying which exact Raman mode is being considered, as different bands can be mediated by different excitons [1,2].

**Intermediate/strong coupling as the cause of observed TERS signal behavior.**

Trying to understand the physics behind the observed behavior, we noticed a fundamental similarity of strongly decreased intensity of the TERS signal from the monolayer $WS_2$ and $MoS_2$ at excitation corresponding to A exciton to the appearance of the so-called Fano/Rabi transparency in scattering/ photoluminescent spectra of TMDs [34–37] or quantum dots[38] in an optical nanocavity formed by a plasmonic nanoparticle and a plasmonic substrate.

It's reasonable to assume that the optical nanocavity volume in our TERS experiments when the metallic TERS probe is in direct contact with the monolayer $WS_2/MoS_2$ which in its turn is in direct contact with the silver substrate, is at least equal if not smaller than the plasmonic nanocavity constructs reported in [34–38], which gives us the reason to state that in our experiments we should implement the conditions for at least the intermediate (Fano resonance) coupling between the TMD excitons and the nanocavity plasmons. Such a coupling explains well a much sharper drop of the monolayer $WS_2$ TERS signal as compared to $MoS_2$, as the A exciton in $WS_2$ is much sharper and further separated from B exciton [39] than in $MoS_2$. Minor variations in the exact position of the A exciton can be a very logical explanation of the variation of the exact shape of the curve of $A'/A_{1g}$ ratio as the function of the excitation laser energy in different crystal tried in our study.

Though the strong coupling between the TMD and the tip-substrate gap plasmons has been demonstrated in TEPL experiments[40], the varied excitation wavelength TERS provides much simpler access to the physics of intermediate- to strongly coupled systems in terms of sample preparation. For TEPL measurements, a dielectric layer separating the TMD or the quantum dots from the metallic plasmonic substrate is a must, yet this is not strictly required for TERS. Decreased optical cavity volume in TERS experiments should enable stronger coupling strength than in case of the sample constructs that preserve the possibility of observation of photoluminescence.

## CONCLUSIONS

In summary, we conducted a systematic study of the dependence of gap-mode TERS spectra of $WS_2$ and $MoS_2$ mono- and bi-layers in intimate contact with silver as a function of the excitation laser wavelength in 830 nm-473 nm range. To achieve this goal, TERS maps of the same area in the sample were collected with the same TERS probe with different excitation lasers. In order to decrease the number of TERS maps collected, we used concurrent excitation/collection with 632.8 nm -785 nm and 532 nm-594 nm laser pairs for $WS_2$, and 632.8 nm-594 nm laser pair for $MoS_2$. To the best of our knowledge, such multi-color TERS imaging is reported here for the first time.

In bilayer $WS_2$@Ag, the $E_{2g}/A_{1g}$ peak height ratio changed in a significantly non- monotonous fashion as a function of the excitation laser energy, having local maxima at energies corresponding to A and B excitons in the bilayer. The A'/$A_{1g}$ peak height ratio went significantly down at an energy corresponding to the A exciton and had an overall descending trend with the increase of the excitation laser energy. A somewhat similar situation was observed in $MoS_2$@Ag where the A'/$A_{1g}$ ratio was highest at 830 nm excitation.

We attribute such behavior to intermediate/ strong coupling between the TMD excitons and the tip-substrate gap plasmons, which on the one hand demonstrates the suitability of simple gap-mode TERS for studies of strongly coupled systems with greater experimental accessibility as compared to TEPL measurements, and on the other hand provides a straightforward rational ( though somewhat counterintuitive) basis for TERS practitioners for the choice of excitation lasers for the study of 2D semiconductors.

Finally, based on experimental observations done in this work and the qualitative explanations of their nature, we proposed an "ideal" gap mode TERS substrate, namely a monolayer h-BN-capped gold or silver which should preserve both the TEPL and TERS signals leaving access to the intermediate-to-strong coupling regime in plasmonic nanocavities.

## ACKNOWLEDGEMENTS

The authors express their deepest gratitude to Dr. Patrick El-Khoury for extremely useful suggestions and guidance provided in the course of this work, as well as to Dr. Gang-Yu Liu of U. C. Davis for precious support with sample preparation. The Stanford work was supported in part by TSMC under the Stanford SystemX Alliance, and by the SUPREME JUMP Center, a Semiconductor Research Corporation (SRC) program sponsored by DARPA. This work was completed in part at the Stanford Nanofabrication and Stanford Nano Shared Facilities, which receive funding from the National Science Foundation (NSF) as part of the National Nanotechnology Coordinated Infrastructure (NNCI) Award ECCS-1542152. J.A.Y. and K.N. acknowledge support from the NSF Graduate Research Fellowship. O.B. acknowledges support from the National Science Foundation (NSF) under the grant numbers, NSF DMR-2117727, and NSF DMR-1720139 for the Materials Research Science and Engineering Center (MRSEC) of Northwestern University. C.F.W. was supported by the United States Department of Energy, Office of Science, Office of Biological and Environmental Research, through the bioimaging technology development program.

## CONFLICTS OF INTEREST

HORIBA Scientific is the manufacturer of the equipment used in this study. Collaboration with industry and academia is a part of A.K. job responsibilities. The authors declare no additional conflicts of interest.

# Excitation laser energy dependence of the gap-mode TERS spectra of WS$_2$ and MoS$_2$ on silver.

# Supporting Information.

Andrey Krayev[1*#], Eleonora Isotta[2*#], Lauren Hoang[3], Jerry A. Yang[3], Kathryn Neilson[3], Minyuan Wang[4], Noah Haughn[4], Eric Pop[3,5], Andrew Mannix[5], Oluwaseyi Balogun[6,7] and Chih-Feng Wang[8].

1. HORIBA Scientific
2. Department of Materials Science and Engineering, Northwestern University
3. Department of Electrical Engineering, Stanford University
4. UC Davis, Department of Chemistry
5. Department of Materials Science and Engineering, Stanford University
6. Department of Mechanical Engineering, Northwestern University
7. Department of Civil and Environmental Engineering, Northwestern University
8. Pacific Northwest National Laboratory

*-Authors contributed equally
#-Corresponding authors

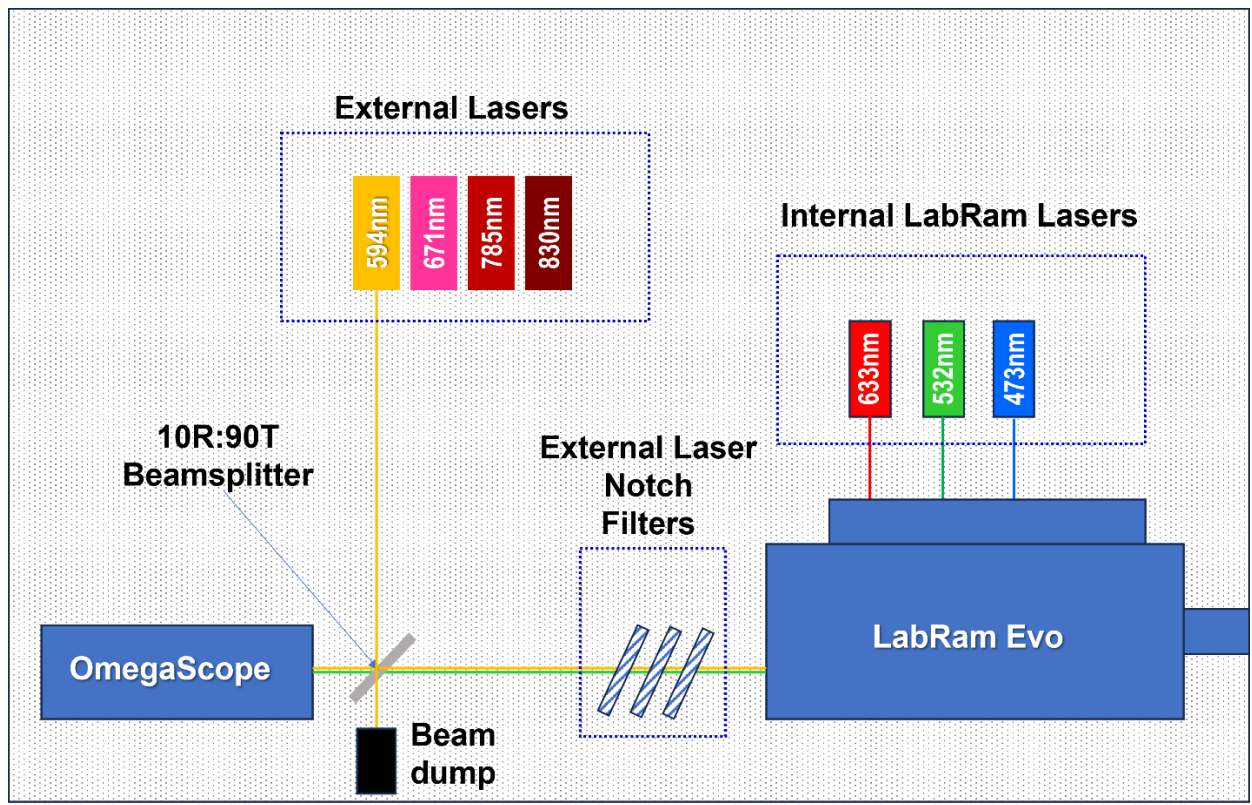

**Figure S1.** Optical scheme of the experimental setup that allows the two-laser concurrent excitation-collection Raman and TERS measurements. The external lasers (594, 671, 785 and

830nm) are injected using 10R:90T beamsplitter and are directed to the AFM part (OmegaScope) along the same path as the internal lasers ( 473, 532 and 633nm) of LabRam EVO Raman spectrometer. External laser excitation laser is filtered out using either ULF notch filters ( for 594nm and 785nm excitation) or regular edge filters ( for 671 and 830 nm lasers). With such a setup and 150 grooves/mm grating the following laser pairs can be used concurrently: 594nm/532nm, 594nm/633nm, 633/785nm.

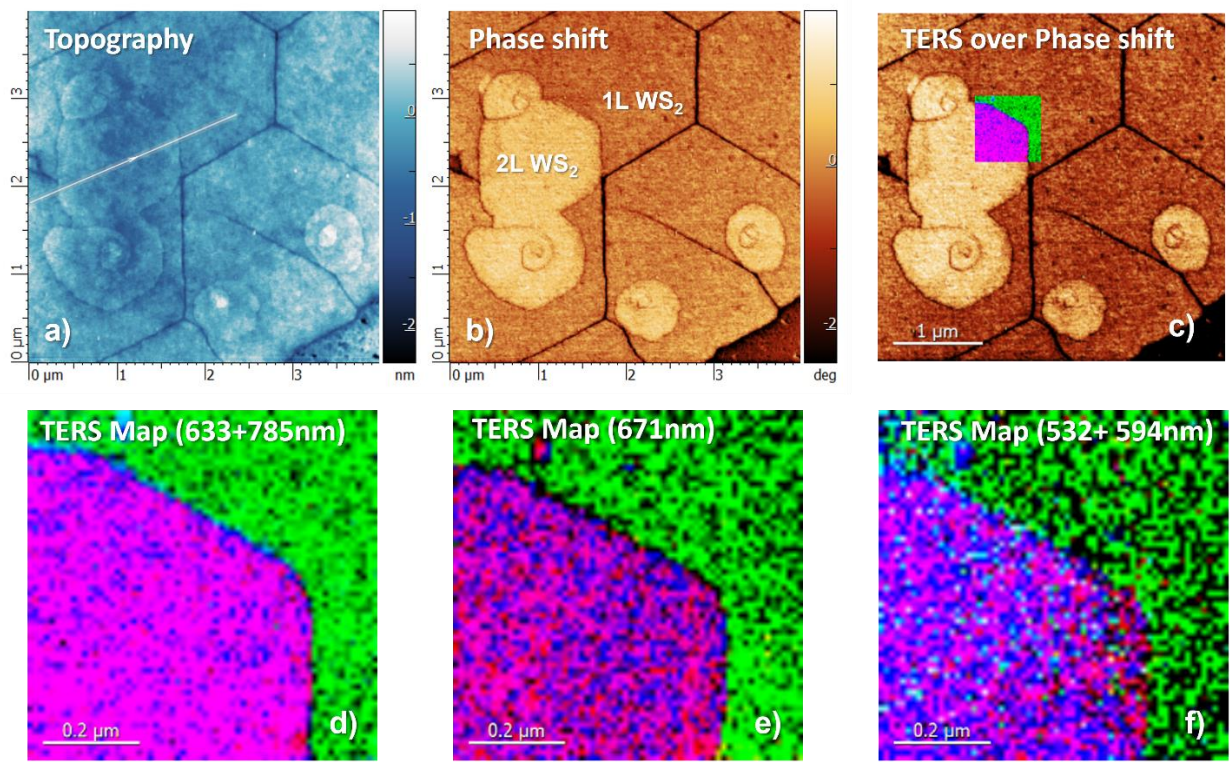

Figure S2. a)- topography and b)- phase shift images of a monolayer $WS_2$ crystal #1 on silver with a bilayer island which is clearly seen both in the topography and in the phase shift images. c)- TERS map overlaid over the phase shift image. d)-, e)-, f)- TERS maps of the same area collected with the same TERS probe using concurrent 633+785nm, 671nm and concurrent 532+594nm excitation.

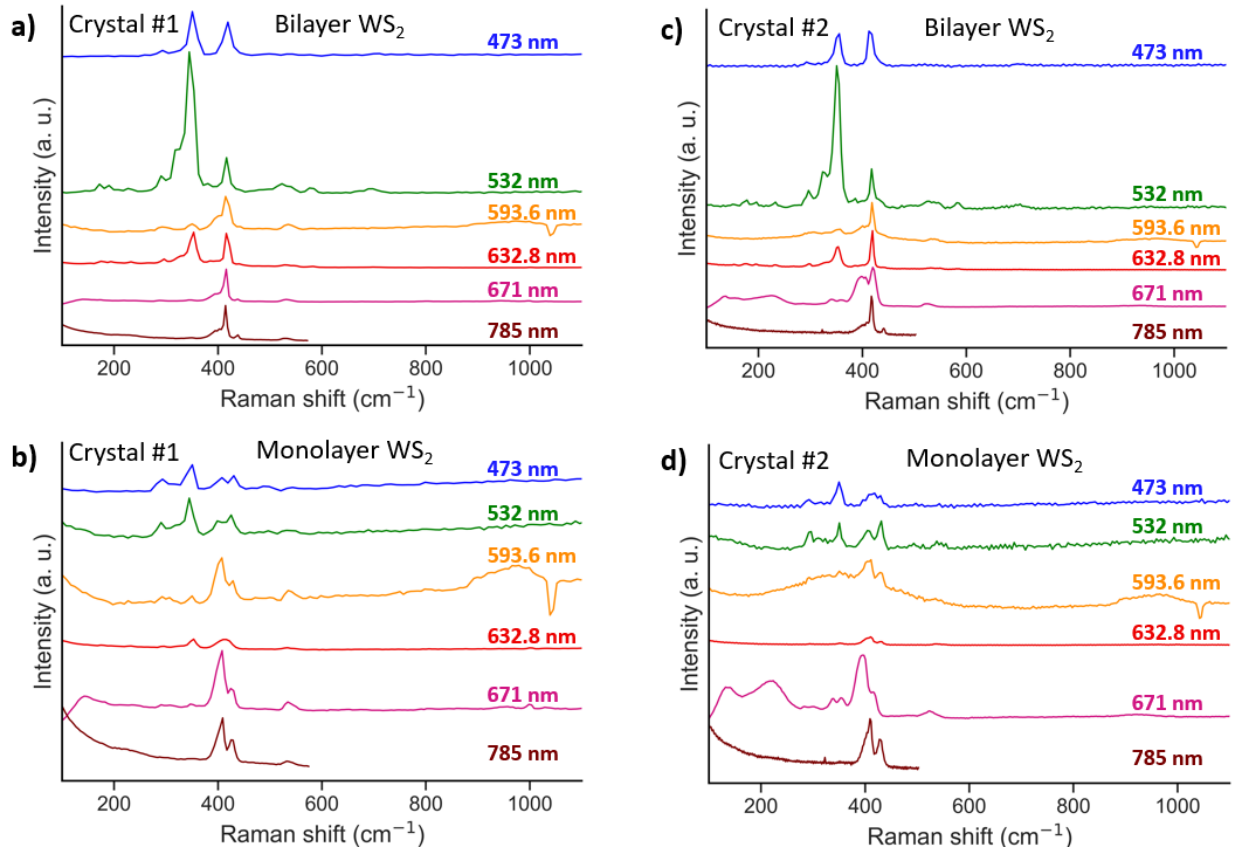

**Figure S3. TERS spectra averaged over a bilayer WS$_2$@Ag for Crystal #1 (a) and Crystal #2 (c) with different excitation lasers and normalized by the height of the A$_{1g}$ peak. Spectra were vertically offset for clarity; TERS spectra averaged over the monolayer WS$_2$@Ag for Crystal #1 (b) and Crystal #2 (d) normalized with the same coefficients as the bilayer spectra in panels a) and c), respectively.**

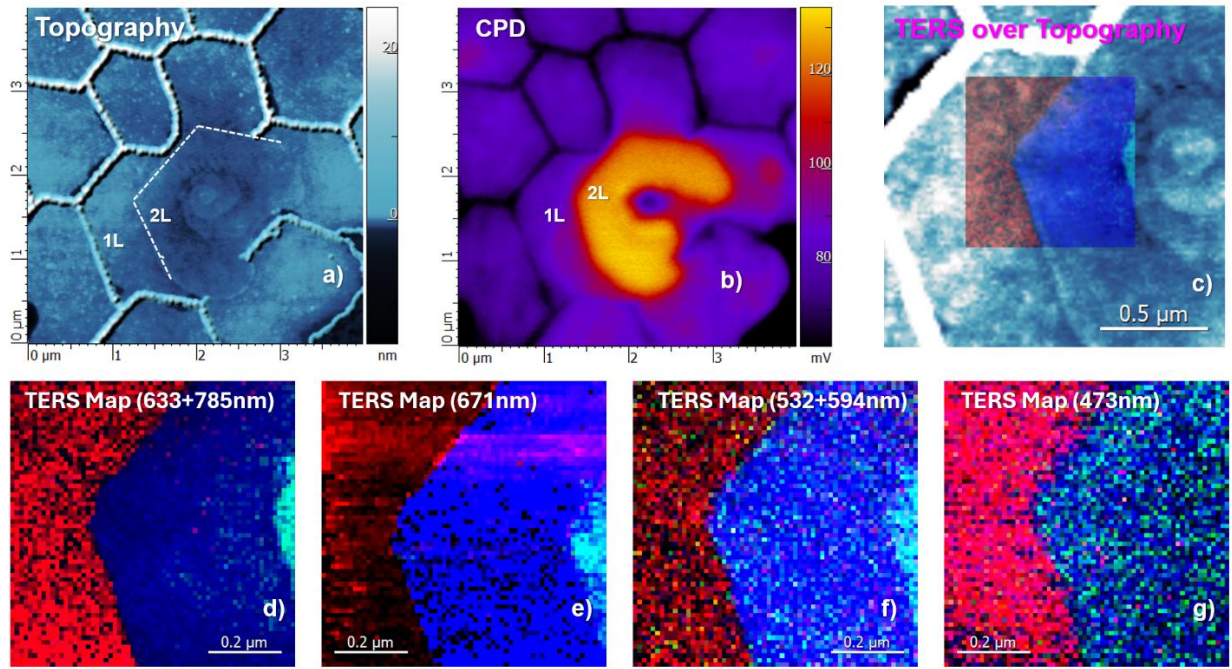

Figure S4. a)- topography and b)- contact potential difference (CPD) images of a monolayer WS$_2$ crystal #2 on silver with a bilayer island which is clearly seen both in the topography and in the CPD images. c)- TERS map overlaid over the topography image. d)-, e)-, f)-, g)- TERS maps of the same area collected with the same TERS probe using concurrent 633+785nm, 671nm, concurrent 532+594nm, and 473nm excitation.